# Galileo (1564-1642) and Kepler (1571-1630): the modern scientist and the mystic[1]

Ivan Todorov

*Institute for Nuclear Research and Nuclear Energy, Bulgarian Academy of Sciences*
*ivbortodorov@gmail.com*

Galileo's place in the history of science has been badly distorted by hero worship.

It was first from his letter to Kepler of 1597, written after reading the preface to the *Mysterium Cosmographicum* which the 26-year old teacher in Graz presented to the professor in Padua, that we learn that "many years ago" Galileo "became a convert to the opinions of Copernicus." Only after 13 more years and the Dutch invention of the telescope, mastered by Galileo, did he make public his views in the *Sidereus Nuncius*, supporting them by his discovery of the Jupiter's satellites, named *Medicea Sidera*. A gifted writer and a brilliant polemicist, Galileo excels in advertising his discoveries - and in making enemies. All that, including his glamorous affair, was instrumental in changing the prevalent philosophy in the "century of geniuses".

Galileo, repelled by Kepler's mysticism (and by his Latin), never read his work. He did not accept Kepler's ellipses (*Astronomia Nova*, 1609) even though Cesi wrote him about them in 1612. They reminded Galileo the deformations of the mannerism paintings of his time which he abhorred. The famous *Dialog* of 1632 never mentions Kepler's laws (or Brahe's observations). The true scientific impact of Galileo comes from his often neglected, 45 years long, "early period" (before *The Starry Messenger*) - in his evolving ideas on motion with Archimedes as spiritual guide: from the balance and the lever, through the pendulum and the inclined plane, towards the law of inertia and the principles of mechanics, eventually published in his *Mathematical Discourses Concerning Two New Sciences* (Elzevir, Leyden, Holland, 1638).

The art of advertising one's scientific achievements, of which Galileo was an early master, is a trademark of successful modern science. Dedicated believers and mystics of science, such as Kepler, are less popular. Yet, an alleged rigorous rationalist like Wolfgang Pauli found in his later troubled life a kinship to Kepler's "archetypal ideas".

## 1. Introduction: "On the shoulders of giants"

Should the phrase[2] "standing on the shoulders of giants", used by Newton in his 1676 letter to Hook (and featuring in the title of Hawking's book[3]), be given a true meaning in the context of Newton's great achievements, it must refer first and foremost to Kepler and Galileo. There is a striking disparity in their public images. In his

---

[1] Outgrow of a conversation over a beer with two Bulgarian poets. Presented to the Third Congress of Bulgarian Physicists.
[2] attributed (by John of Salisbury in 1159) to Bernard of Chartres of 12th century.
[3] *On the Shoulders of Giants, The Great Works of Physics and Astronomy,* edited, with commentary.

Introduction to [C] Owen Gingerich, Harvard's astronomer and historian of science, recalls the motivation of [K][4]: While Galileo and Kepler were the two giants on whose shoulders Newton had stood, why was the name of the first familiar to every schoolboy, but the second known to only a small number of intellectuals? At closer reading popularity appears, as usual, intertwined with legend. Galileo's legend is related to the glamorous trial of 1633 which is often presented as typifying the clash between modern science and the Catholic Church - a crude simplification, to say the least. Contrary to traditonal accounts, we are jumping over this popular story altogether, paying more attention instead to some of the great achievements of the two men and to their contrasting approach to science. Readers interested in the ideological struggle(s) of Galileo will find a careful analysis in the corresponding articles of [CC], and, may be most thoroughly, in the recent Russian language monograph [D].

## 2. The Renaissance man of Tuscany and the Swabian mystic

After the "classic work dealing with Galileo's life and scientific achievements" [D78], I enjoyed reading the more recent and lighthearted "magnificent biography" (in the words of Peter Machamer [M]) [H] and will try to share some of its flavor:

"Although Galileo was born in Pisa (in 1564), the hometown of his recalcitrant mother, he prided himself on being a noble of Florence through his father, Vincenzo Galilei, a musician and musical theorist." (p. 2 of [H]). Galileo's characteristic in the Preface of [H] is particularly eloquent: "Galileo enjoyed such epithets as "divine mathematician" and "Tuscan Archimedes," and he spent the first half of his career, from 1589 to 1610, as a professor of mathematics. ... For all that, he was no more (or less!) a mathematician than he was a musician, artist, writer, philosopher, or gadgeteer. His last disciple and first biographer, Vincenzo Viviani, boasted that his master could compete with the best lutanists in Tuscany, advise painters and poets on matters of artistic taste, and recite vast stretches of Petrarch, Dante, and Ariosto by heart. But his great strength, Galileo said when negotiating for a post at the Medici court in 1610, was philosophy, on which he had spent more years of study than he had months on mathematics... Galileo the patrician humanist ... underwent a sort of epiphany under the impetus of the telescopic discoveries he made at the age of 45. He had published very little, and nothing of importance, up to that time. He had many good ideas, but held them back...When he had armed himself with the telescope, however, he declared all he knew and more. To the surprise of his colleagues and against their advice, he attacked philosophers, theologians, and mathematicians, taunted the Jesuits, jousted with everyone who contested his priority or his opinions. He became a knight errant, quixotic and fearless, like one of the paladins in his favorite poem, Ariosto's Orlando furioso. This change in behavior, which won him a continually lengthening list of enemies, made his disastrous collision with a pope who for many years had been his friend and admirer intelligible and even inevitable."

Kepler's biographies are much fewer (than Galileo's) and are mostly based on

---

4  whose central piece, *The Watershed,* is the first notable English language biography of Kepler.

[C]. From the introduction to the Dover edition of [C]: "Caspar was eminently qualified to write the standard biography. Like Kepler himself, Caspar was born in southern Germany, had been trained in both theology and mathematics at Tübingen..."

At the age of 25, Kepler (born in Weil der Stadt, "gate to the Black Forest", in 1571) drew unflattering portraits of his parents and ancestors comparing them with their horoscopes ([C] Sect. I.3). He remembered, though, how his mother showed him the great comet of 1577. Matriculated at the University of Tübingen in 1587, he was influenced by Maestlin, his astronomy professor, who knew Copernican astronomy well (his "1543 *De revolutionibus* is probably the most thoroughly annotated copy extant" [G]). Leaving, against his will, the hope to become a clergyman in Tübingen, Kepler found his true calling as a "theologian cosmologist". On the eve of publishing his first book, the *Mysterium cosmographicum* of 1596, the first unabashedly Copernican treatise since *De revolutionibus* itself, he wrote to his teacher (Maestlin): "I wanted to become a theologian, for a long time I was restless. Now, however, behold how through my effort God is being celebrated in astronomy." [G]. This work was for him the beginning of a big project that included [Astronomia nova](), [Harmonices Mundi](), and [Epitome of Copernican Astronomy]() where his three famous laws were formulated. To quote [G]: *Kepler's scientific thought was characterized by his profound sense of order and harmony, which was intimately linked with his theological view of God the Creator. He saw in the visible universe the symbolic image of the Trinity. Repeatedly, he stated that geometry and quantity are coeternal with God and that mankind shares in them because man is created in the image of God*...[5] Kepler wrote prolifically, but his intensely personal cosmology was not very appealing to the rationalists of the generations that followed. A much greater audience awaited a more gifted polemicist, Galileo, who became the persuasive purveyor of the new cosmology. *Kepler was an astronomer's astronomer*. It was the astronomers who recognized the immense superiority of the *Tabulae Rudolphinae*.

The nature of Kepler's religious views and their unifying role in his work were analyzed by Holton [H56]: From his earliest writing to his last, Kepler maintained the direction and intensity of his religio-philosophical interest... Next to the Lutheran God, revealed to him directly in the words of the Bible, there stands the Pythagorean God, embodied in the immediacy of the observable nature and in the mathematical harmonies of the solar system whose design Kepler himself had traced - God "whom in the contemplation of the universe I can grasp, as it were, with my very hands." (letter to Baron Strahlendorf, October 1613). Or, in an early letter to his teacher: "the belief in the creation of the world be fortified through this external support, that the thought of the creator be recognized in its nature ... Then man will at last measure the power of his mind on the true scale, and will realize that God who founded everything in the world according to the norm of quantity, also has endowed man with a mind which can comprehend these norms. For as the eye for the color, the ear for the musical sound, so is the mind of man created for the perception not of any arbitrary entities, but rather of quantities; the mind comprehends a thing the more correctly the closer the thing approaches toward a pure quantity as its origin." (letter to Maestlin, April 1597) ... Kepler saw ... the universe as a physical machine, ... as mathematical harmony, and ... as central theological order. And this was the setting in which conception of the universe led to specific results of crucial importance.

---

5   These words (as well as those in the letter to Maestlin below) are echoed by Salviati (of Galileo's dialogue of 1632) who elevates true achievements of science to "those few which the human intellect does understand, I believe that its knowledge equals the Divine in objective certainty" [Ga].

### 3. Early period. First exchange. **The 8-minutes error and the ellipses**

The long neglected Galileo's "early period" (the first 45 years (!) of his life) is important both for displaying his dept to teachers, predecessors (such as Borro, [H], Sect. 2.3, Benedetti, [Koy], 140-165) and colleagues and for revealing the difficulties he had to overcome on his road to the law of inertia. To quote Hooper [CC]: Classical mechanics is still taught by referring new students to the core set of problems that had to be solved by the original investigators like Descartes, Gassendi, Huygens, Wallis, Wren, Hooke, and Newton, all following Galileo's original line of attack. These problems include the analysis of motion on an inclined plane, the motion of a pendulum, the action of a lever, the force of a spring or pull in a rope, the result of collisions between impacting and moving bodies, and so on. The difficulty with the law of inertia stems from the fact that it is never valid on earth because of gravity which was only understood later, in the work of Newton. Galileo analyzed projectile motion into two component motions, the first horizontal and uniform, the other vertical and accelerated. Galileo discussed the motions of bodies upon the moving Earth and of planets around the Sun. He asked questions that led his fellows and successors directly toward inertial mechanics and gave them some of the essential tools to build it.

Galieo was teaching (in Pisa and then in Padua) Ptolemy, preferring privately Copernicus as witnessed by a long letter of 1597 to his elder Pisan friend Mazzoni ([H], *The Copernican confession*). A few months later Galileo received from the hands of a personal messenger a copy of young Kepler's *Mysterium cosmographicum.* Like many people who receive unexpected books, Galileo thanked the author immediately so as not to have to comment in detail. He had had time only to read the preface, he said, from which he gathered that congratulations were in order, not to the writer, but to the reader, for "having acquired such a lover of truth as an ally in the search for truth." Kepler had found some choice things, which Galileo promised to study, "and that the more willingly since I adopted Copernicus' opinion many years ago, and deduced from it the causes of many natural effects doubtless inexplicable on the ordinary hypothesis. I've written out many reasons for it and many responses to reasons against it, which I have not dared to publish as I've been deterred by the fate of our master Copernicus. For although he has gained immortal fame among a few, he has been ridiculed and derided by countless others (for such is the number of fools). I would venture to disclose my thoughts if there were more like you; but as there are not, I will forbear." Kepler tried to stiffen the backbone of his shy ally. "I was very pleased to receive yours of 4 August, firstly because of friendship begun with an Italian and secondly because of our agreement about Copernican cosmology." Mathematicians everywhere (Kepler continued) side with Copernicus and calculate according to his principles. If we all speak out together, people ignorant of mathematics will have to take our word for it. "If I'm right, not many good mathematicians in Europe will wish to differ from us; tanta vis est veritas, such is the power of truth. If Italy is not a suitable place for publication, and if you encounter other difficulties, perhaps Germany will grant us this freedom . . . Have faith, Galileo, and go forth." To this pep talk, and an appended request to make a certain astronomical observation in the common cause, Galileo did not respond at all.

     A teacher at the Lutheran school in Graz (asked to teach Virgil, rhetoric and arithmetic) young Kepler made his mark by issuing a calendar and prognostication for 1595, which contained predictions of bitter cold, peasant uprisings, and invasions by the Turks. (All were fulfilled, to the great enhancement of his local reputation.) Meanwhile, just over a year after his arrival in Graz, Kepler's fertile imagination hit upon what he believed to be the secret key to the universe. There were six known planets at the time and there are exactly five regular polyhedrons (*Platonic solids*: the tetrahedron, cube and octahedron, dodecahedron and icosahedron). Kepler devised a scheme (that worked fairly well [G]) in which each planet moves on a circle inscribed or superscribed around corresponding Platonic solids. Although *Mysterium cosmographicum* was leading in a wrong direction, Kepler established himself as the first, and until Descartes the only, scientist to demand physical explanations for celestial phenomena. Seldom in history has so wrong a book been so seminal in the future course of science.

     Providence kept helping Kepler as if against his will: By the fall of 1598 Catholic rulers in Graz started chasing away protestants. Being not welcome at his Alma mater (in Tübingen) he had to go to Prague where, upon the death of his host, the great Danish astronomer Tycho Brahe (1546-1601), he became imperial mathematician in the court of Rudolph II. Luckily, he had been assigned (by Tycho) to study the orbit of Mars, the planet with greatest eccentricity, which helped him liberate astronomy from the two-thousand-year-old dogma of circular motion. Remarkably, explaining the precise observations of Tycho was more important to Kepler than a priory aesthetic ideas: "Divine Providence granted us such a diligent observer in Tycho Brahe," he wrote, "that his observations convicted this Ptolemaic calculation of an error of 8'; it is only right that we should accept God's gift with a grateful mind.. . . Because these 8' could not be ignored, they along have led to a total reformation of astronomy." The first two laws were thus mastered essentially already in *Astronomia nova* (1609) but the precise formulation of all three planetary laws only appears in book V of his *Epitome astronomiae Copernicanae* (1621). The puzzling fact that Galileo never took seriously Kepler's ellipses is explained in [P56] by his aesthetic views: for Galileo the ellipse is a deformed circle reminding him the deformed human faces in the then becoming fashionable mannerism paintings (an opinion also supported in Koyré's *Attitude esthétique et pensée scientifique*, [Koy], pp. 275-288).

## 4. The Starry Messenger. Theories of tide

     During his Paduan tenure Galileo befriended the enlightened Copernican and

influential Venetian Sarpi[6]. In the summer of 1609 a claim came to Italy of Dutch spectacle makers to a gadget that made distant objects appear near. One came into Sarpi's hands in July 1609. Having examined it, he could advise the Senate not to buy it from a traveling salesman who had offered it, together with its "secret," for 1,000 scudi. By then, August 1609, the secret was out. Sarpi's knowledge of optics gave him confidence that the gadget could easily be bettered, and his knowledge of men assured him that Galileo was the one for the job. As Sarpi wrote to a friend, *The Dutch gadget became the Italian telescope* through the efforts of "the mathematician [Galileo] and others here [in Venice] not ignorant of these arts."[H]. In December 1609 Galileo raised his best telescope, then of 20x, to the sky, an exercise for which he was fully prepared (with his firsthand knowledge of perspective among other things). Sometime before 7 January 1610, when Galileo described his lunar discoveries to Antonio de' Medici, he noticed through his 20x telescope that Jupiter had lined up along the ecliptic with three little stars. Galileo immediately recognized a life chance for a real discovery. Even if a friend first saw the event, as the jealous successor of the Florentine in Padua had it [H], Galileo alone was able to identify Jupiter's starlets as elements of a miniature solar system. That took immense skill and application; or "the carefulness and industry of a Florentine." One can follow this care and industry day by day in Galileo's drawings of the changing configurations of Jupiter and the starlets. Galileo's account of his discoveries, rushed into print early in March 1610 under the title *Sidereus nuncius*, included the fanciful designation of Jupiter's moons as Medici stars. Galileo's discoveries were met with skepticism and mistrust, especially in his native Italy; so in April 1610, he sent his book to Kepler in Prague, requesting an opinion. Kepler's response was enthusiastic and generous. Even before having observed Jupiter's moons himself, he starts his message - *Dissertatio* with: "Whom does knowledge of such important things allow to be silent?" ([C] Sect. III.14, p. 192).

      A few remarks are in order. - Galileo never mentioned his human debt to friends and colleagues in Venice and Padua (neglecting to consider the importance of the testimony of trustworthy Venetians, able to certify that the discoveries announced were not optical illusions): he was preoccupied with flattering his former pupil Cosimo II de' Medici while negotiating best possible conditions for his tenure at the Tuscan court. -- He wrote the *Starry message* in Latin, as befitted to a scientific discovery, Galileo's most important contribution to the field of astronomy. (By contrast, his famous *Dialogue Concerning the Two Chief World Systems* of 1632, a masterpiece of Italian prose, is a speculative polemical exposé of 16th century Copernican physics that ignores newer observations and theoretical development by Tycho Brache and Kepler.)

      It is interesting to compare the different approaches of Galileo and Kepler to similar problems. When Kepler has to face optical observation he studies the theory - in *Astronomiae pars optica* (1604), ..., *Dioptrice* (1611), founding on the way the geometric optics. Galileo is playing instead with two lenses and soon produces an improved telescope. Kepler is spending years searching for "the third law of planetary motion" - the precise relation between the cubes of large semiaxes and the squares of

---

6   **Paolo Sarpi** (1552 – 1623) was an Italian historian, scientist, statesman, active on behalf of the Venetian Republic during the period of its successful defiance of the papal interdict (1605–1607).

the corresponding periods. Galileo collects similar data for Jupiter's satellites but does not look for a relation between them thus missing the opportunity to be the first to discover the third law. For him "mathematics" is the Archimedian geometry: he has no taste for analytic and algebraic computations.

Perhaps the most instructive example of a clash between Galileo's smooth "rational thinking" and Kepler's "mysticism" is provided by their different approaches to the theory of tides. In 1616 Galileo published (in Italian) his *Discorso* on the topic. In his view, it provided *The decisive proof that the Earth moves* [S], p. 224 (the idea having come to him in a flash on one of his frequent trips from Padua to Venice in a large barge whose bottom contained a certain amount of water). Kepler had the right intuition that the tides are caused by the moon's attraction - a view confirmed and further elaborated by Newton and Laplace of the next generations. To quote [H], Sect. 7.2, p. 260: *Galileo's prevailing misjudgments as a natural philosopher come into view here. Neglecting physical cause, he advanced his pendulum analogy, which was no more than a metaphor, as an explanation. What is it that binds the earth and moon so strongly together that they act as a single pendulum bob? Galileo liked the analogy all the more for this weakness. In the paradoxical way he loved, it gave the moon a role in the drama of the tides "without [its] having anything to do with oceans and with waters." It also allowed him to sidestep the hidden connection between the lunar motions and the diurnal tides, and to rap Kepler, who, "though he had at his fingertips the motions attributed to the earth . . . has nevertheless leant his assent to the moon's dominion over the waters."* In fact, Kepler has anticipated the law of universal gravitation. He stated that gravity was a *mutual* tendency between material bodies toward contact, so the earth draws a stone much more than the stone draws the earth. Heavy bodies are attracted by the earth not because it is the center of the universe, but simply because it contains a lot of material, all of which attracts the heavy body. Kepler realized that the tides were caused by the waters of the oceans being attracted by the moon's gravitational pull. He wrote (in the Introduction to *Astronomia Nova*): *"If the earth ceased to attract the waters of the sea, the seas would rise and flow into the moon..."* and went on to add: *"If the attractive force of the moon reaches down to the earth, it follows that the attractive force of the earth, all the more, extends to the moon and even farther..."* (We recommend the well documented emotional expossition of Sects. 6.8-10, pp. 334-343, of Koestler's book [K] where these quotations are put into context.) One should be also able to understand why for Galileo the mutual attraction at a distance of celestial bodies sounds like a magic. Even Newton has expressed his dissatisfaction in his philosophical queries (if not in the *Principia*). Only with the advent of general relativity one begins to understand gravitation as a local field theory: a dynamical change of space-time geometry by moving bodies.

Quite apart from the theory of tides, this is a good place to illustrate why does one need the insight of both Kepler and Galileo for the Newton synthesis. It seems incredible, with hindsight, that Kepler could have understood the gravitational force so well, and yet it did not apparently occur to him that it might play a central role in determining the orbital motions of the planets! The essential reason he failed to make the connection was that he had no intuition for the inertial movement: *he believed the* *planets needed a constant pushing force, in the direction of motion, to keep them going* in their orbits. This was an ancient belief that Galileo demolished in his discussions of projectiles in *Discourses on the Two New Sciences* (1638). But albeit the Discourses resurrected some work of his "early period" it was only published after Kepler's death (1630). Galileo's insight about projectiles was then extended to the planets by Newton.

### 5. Final years. Kepler's wine barrels and Galileo's Tuscan wine

When the deposed emperor Rudolph died in January 1612 Kepler went to Linz as provincial mathematician, a post created specially for him. Although his most creative period was laying behind him, his fourteen-year sojourn in Linz eventually saw the production of his *Harmonice mundi* and *Epitome astronomiae Copernicanae* and the preparation of the *Tabulae Rudolphinae*. One bright spot in his Linz career was his second marriage, to Susanna Reuttinger, a twenty-four-year-old orphan, on 30 October 1613. In an extraordinary letter to an unidentified nobleman, Kepler details his slate of eleven candidates for marriage and explains how God had led him back to number five who had evidently been considered beneath him by his family and friends. The marriage was successful, far happier than the first; but of their seven children, five died in infancy or childhood. Likewise, only two of the five children of his first marriage survived to adulthood.

That Kepler, engulfed in a sea of personal troubles, published no astronomical works from 1612 through 1616 is not surprising. Yet he did produce the *Stereometria doliorum vinariorum* (1615), which is generally regarded as one of the significant works in the prehistory of the calculus. Desiring to outfit his new household with the produce of a particularly good wine harvest, Kepler installed some casks in his house. When he discovered that the wine merchant measured only the diagonal length of the barrels, ignoring their shape, Kepler set about computing their actual volumes. Captivated by the task, he extended it to other shapes, including the torus.

In his own eyes Kepler was a speculative physicist and cosmologist; to his imperial employers he was a mathematician charged with completing Tycho's planetary tables. He spent most of his working years with this task hanging as a burden as well as a challenge; ultimately it provided the chief vehicle for the recognition of his astronomical accomplishments. In excusing the long delay in publication, which finally took place in 1627, he mentioned in the preface not only the difficulties of obtaining his salary and of the wartime conditions but also "the novelty of my discoveries and the unexpected transfer of the whole of astronomy from fictitious circles to natural causes, which were most profound to investigate, difficult to explain, and difficult to calculate, since mine was the first attempt." Kepler realized that the improved accuracy of his tables enabled him to predict a pair of remarkable transits of Mercury and of Venus across the disk of the sun. These he announced in a small pamphlet, *De raris mirisque anni* 1631 *phenomenis* (1629). Although he did not live to see his predictions fulfilled, the Mercury transit was observed by Pierre Gassendi in Paris on 7 November 1631.

The 58-year old Kepler died in Regensburg on November 15, 1630 while traveling to collect his salary. He was buried in the Protestant cemetery; the churchyard was completely demolished during the thirty years war. Jacob Bartsch, who had married Kepler's daughter Susanna in March 1630, became a faithful protector of the bereaved and penniless family. He recorded the epitaph that Kepler himself has composed: *I used to measure the heavens, now I shall measure the shadows of the earth...*

The final period of Galileo's life, starting with his *Dialogue Concerning the Two Chief World Systems,* falls after Kepler's death. Picking himself up from his humiliating posture before the cardinals and the gospels, Galileo received permission to stay within the palace of the archbishop of Siena, Ascanio Piccolomini, in anticipation of a return back home to Arcetri after an absence of over a year. The six months that Galileo spent in Siena at Piccolomini's house and table revived his spirits[7]. He started a new work on mechanics—"full of many curious and useful ideas" - resurging his youthful thoughts. Galileo was enjoying premium wine at the archbishop's table (not trying to determine the volume of the casks) as can be surmised from the letters of his loving daughter Maria Celeste[8]: "I pray that you continue [in good health] by governing yourself well particularly with regard to the drinking that is so hurtful to you ...". Thus Galileo, like Kepler, completes and publishes his ripest work *Mathematical Discourses Concerning Two New Sciences*, which crowns his oeuvre, during the last years of his life. It is a no small feat for the embittered blind old man who has just lost his favorite child. The book resumes the discussions of the three participants of the condemned *Dialogue.* They no longer mention Copernicus but do praise "our Academician" (i.e. Galileo) giving an occasion to the witty Descartes (who never masters Galileo's law of free falling bodies - see [D], Part II) to ironize: *"[Galileo's] way of writing in dialogues with three persons who do nothing but praise and exalt his inventions in turn certainly makes the most of his wares."*

## 6. Epilogue: Pauli's Kepler

*I never understood what Jung did to Pauli*.

(from a letter of a colleague)

In response to suggestions to publish his article on Kepler separately from Jung's essay on synchronicity Wolfgang Pauli (1900-1958) told his former assistant Markus Fierz (1912-2006): "I have thought about it and I believe I should not do this. For, indeed, there comes the time when I must give documentary evidence of what I owe this man" (quoted by Fierz in 1957 - see [P] p. 5). But Pauli's interest in Kepler and his "mystical reflections" predated his encounter with Jung. To cite [M09] ("the first popular biography" of Pauli according to G. Farmelo): "Since his student days and before, Pauli had been interested not just in the rational world of physics but also in the role of the irrational. Arnold Sommerfeld, his professor and lifelong mentor, ... reminded his students, that science emerged out of mysticism and had never completely separated itself. Besides his purely scientific work, Sommerfeld also pursued kabbalistic lines of research based on pure numbers and spoke of Kepler as his precursor" (see Chapter 5: *Intermezzo - Three versus Four. Alchemy, Mysticism and the Dawn of Modern Science*). Here is how Pauli himself characterizes his teacher (in an article of 1948, dedicated to Sommerfeld's 80th birthday (see [P] 5.): "In the Preface to the first edition [of his *Atombau und Spektrallinien*] he conjured up, along with Kepler's ellipses, the spirit of Kepler as well, when he wrote: *What we are nowadays hearing of the language of spectra is a true music of the spheres within the atom, chords of integral relationships, an order and harmony that becomes ever more perfect in spite of the manifold variety... It is the mysterious organon on which Nature plays her music of the spectra, and according to the rhythm of which she regulates the structure of the atoms and nuclei.* It is though as there were here an echo of Kepler's search for the harmonies in the cosmos, guided by the musical feeling for the beauty of just proportion in the sense of Pythagorean philosophy, - an echo of his "geometria est archetypus pulchritudinis mundi" (geometry is the archetype of the beauty of the universe)."

---

7  The fine story that when rising from his knees before the inquisition Galileo muttered, "still it moves," is associated with the Piccolomini. A portrait, representing the scene of Galileo's recantation, perhaps by Murillo, displays the slogan, *eppur si muove* ([H], p. 327).

8  Galileo's elder daughter Virginia (Suor Maria Celeste, 1600-1634) was confined by him together with her sister to the convent of San Matteo in Arcerti and is buried with her father at Santa Croce - [So].

The truth is that Pauli's personal crisis[9] of 1930, which led him to address the famous psychiatrist in 1931, came in the middle of his life, before any of his public pronouncements in the above spirit. On the other hand, his widow, Franca (1901-1987), "did everything in her power to consign the 'Jungian' part of Pauli's thinking to oblivion" (see [G05], p. 4). Quite a few years after her death had to pass before the Pauli Committee and other experts finally yielded to the argument that 'it is of no importance what we think of Jung and his psychology. The important thing is that Pauli was a convinced adherent of Jung's teachings. One cannot therefore leave out this part of his writing and his estate.' ([G05] p. 5). There is a fallacy even in this admission. As Heisenberg, Pauli's longtime friend, has correctly pointed out [H59] Pauli followed since his early days the skeptical way, based on the usage of reason, reaching the point of exercising skepticism towards the skepticism itself, before trying to tackle the complicated path to knowledge. Then Jung's teachings fell on a fertile ground and were enriched by Pauli.

Pauli's study of the archetypal ideas of Kepler ([P] 16.), published in 1952, was based on his lectures in February and March 1948 at the psychological club in Zürich, but his interest in Kepler is documented since (at least) 1938; it was further stimulated by his discussions with the art historian Erwin Panofsky in Princeton in 1940-1946 and by reading (in 1947) Markus Fierz's essay on Newton (see [G05] pp. 179-183).

Here is a crucial passage in the first section of Pauli's article ([P21] p. 221) which speaks for itself: "The process of understanding nature as well as the happiness that man feels in understanding, that is, in the conscious realization of new knowledge, seems thus to be based on a correspondence, a "matching" of inner images pre-existent in the human psyche with external objects and their behavior. This interpretation of scientific knowledge, of course, goes back to Plato and is, as we shall see, very clearly advocated by Kepler. He speaks in fact of ideas that are pre-existent in the mind of God and were implanted in the soul, the image of God, at the time of creation. These primary images which the soul can perceive with the aid of an innate "instinct" are called by Kepler archetypal ("archetypalis"). Their agreement with the "primordial images" or archetypes introduced into modern psychology by C. G. Jung and functioning as "instincts of imagination" is very extensive." Pauli quotes further Keppler's formulation of the "matching" of external impressions with pre-existing inner images (in his *Harmonices mundi*): *For, to know is to compare that which is externally perceived with inner ideas and to judge that it agrees with them, a process which Proclus expressed very beautifully by the word "awakening", as from sleep.*

---

9   His mother committed suicide in 1927 after his father left her to marry a much younger woman. Next year the 28-year-old Pauli had been given the chair of theoretical physics at ETH, Zürich. In 1929 he left the Catholic Church and in December married a cabaret dancer who left him soon for a chemist...

If it looks strange to a modern reader that a critical rational scientist like Pauli may find a kinship with Kepler's archetypal images, Pauli's attitude towards Kepler's discussion with the Rosicrucian alchemist Robert Fludd will appear even stranger. Let us again give the word to Pauli ([P] 21. p. 257): "In the first half of the seventeenth century when the then new, quantitative, ... mathematical way of thinking collided with the alchemical tradition expressed in qualitative, symbolical pictures: the former represented by the productive, creative Kepler always struggling for new modes of expression, the latter by the epigone Fludd who could not help but feel clearly the threat to his world of mysteries, already become archaic, from the new alliance of empirical induction with mathematically logical thought. One has the impression that Fludd was always in the wrong ... Due to his rejection of the quantitative element he remained unconscious of its laws and inevitably came into ... conflict with scientific thinking.

Fludd's attitude, however, seems to us somewhat easier to understand when it is viewed in the perspective of a more general differentiation between two types of mind, a differentiation that can be traced throughout history, the one type considering the quantitative relations of the parts to be essential, the other the qualitative indivisibility of the whole. We already find this contrast, for example, in antiquity in the two corresponding definitions of beauty: in the one it is the proper agreement of the parts with each other and with the whole, in the other (going back to Plotinus) there is no reference to parts but beauty is the eternal radiance of the "One" shining through the material phenomenon. An analogous contrast can also be found later: Goethe had a similar aversion to "parts" and always emphasized the disturbing influence of instruments on the "natural" phenomena. We should like to advocate the point of view that these controversial attitudes are really illustrations of the psychological contrast between feeling or intuitive type and thinking type. Goethe and Fludd represent the feeling type and the intuitive approach, Newton and Kepler the thinking type; even Plotinus should ... not be called a systematic thinker, in contrast to Aristotle and Plato. "

Pauli attempts to harmonize the two attitude by invoking Bohr's complementarity: "The general problem of the relation between psyche and physis, between the inner and the outer, can, however, hardly be said to have been solved by the concept of "psychophysical parallelism" which was advanced in the last century. Yet modern science may have brought us closer to a more satisfying conception of this relationship by setting up, within the field of physics, the concept of complementarity. It would be most satisfactory of all if physis and psyche could be seen as complementary aspects of the same reality." ([P] 21. p. 260).

Jung's essay on "synchronicity" was rejected by "serious scientists"; Pauli was himself critical to some parts of it "which only encouraged believers in astrology".

Nevertheless he wrote to Fierz (on Christmas) in 1954: "Many physicists and historians have of course advised me to break the connection between my Kepler essay and C. G. Jung in the English translation. On mature reflection however I have still decided not to do so: it is not important to be entirely loyal personally to C. G. Jung (and not so important to him either). But it is very important to remain loyal to my own unconscious. ... I am indifferent to the astral cult of Jung's circle, but that, i.e. this dream symbolism, makes an impact! The book is itself a fateful 'synchronicity' and must remain one. ... I am sure that defiance would have unhappy consequences as far as I am concerned. *Dixi et salvavi animam meam!"* (see [G05] p. 298).

The latest text of Pauli concerning science and Western thought is his talk at a 1955 workshop in Mainz. Here is a passage where he mentions both Galileo and Kepler ([P] 16. p.140): "Among the attempts that have occurred in the course of history to effect a synthesis of the basic attitudes of science and of mysticism there are two which I should like particularly to stress. One of these originates with Pythagoras in the sixth century B. C , is then carried on by his disciples and developed further by Plato, appearing in late antiquity as Neo-Platonism and Neo-Pythagoreanism. Since much of this philosophy was taken over into early Christian theology, it continues thereafter in persevering association with Christianity, to blossom anew in the Renaissance. It was through the rejection of the anima mundi, the world-soul, and a return to Plato's doctrine of knowledge in Galileo's work, and through a partial revival of Pythagorean elements in that of Kepler, that the science of modernity, which we now call classical science, arises in the seventeenth century."

Pauli then returns to the Goethe-Fludd analogy ([P] 16. p. 146): "Goethe's scientific conceptions, which were so often in opposition to official science, become more comprehensible in the light of their alchemical sources, the terminology of which comes to light quite plainly, especially in "Faust". Goethe was an emotional type and hence more susceptible to the experience of unity - "nothing inside, nothing outside, for what's inside that is outside" - than to critical science. In this regard it was alchemy alone that suited his emotional attitude. This is the background of Goethe's antagonism to Newton on which much has been written. Less known are the earlier polemics between Kepler, representing science just developing, and the English physician Robert Fludd, who ... represented the Hermetic tradition. I believe that one is justified in applying to Kepler-Fludd and Newton-Goethe the old saying "Was die Alten sungen, das zwitschern die Jungen" (the young twitter as the old folk sang)."

Pauli ends his essay by invoking again Bohr's complementarity in an attempt to restore the equilibrium between the rational and the mystical still present in Kepler's time: "... at present a point has again been reached at which the rationalist outlook has passed its zenith, and is found to be too narrow. Externally all contrasts appear to be extraordinarily accentuated. On one hand the rational way of thought leads to the assumption of a reality which cannot be directly apprehended by the senses, but which is comprehensible by means of mathematical or other symbols, as for instance the atom or the unconscious. But on the other hand the visible effects of this abstract reality are as concrete as atomic explosions, and are by no means necessarily good, indeed sometimes the extreme opposite. A flight from the merely rational, in which the will to power is never quite absent as a background, to its opposite, for example to a Christian or Buddhist mysticism is obvious and is emotionally understandable. Yet I believe that there is no other course for anyone for whom narrow rationalism has lost its force of conviction, and for whom also the magie of a mystical attitude, experiencing the external world in its crowding multiplicity as illusory, is not effective enough, than to expose hirnself in one way or another to these accentuated contrasts and their conflicts. It is precise1y by this means that the scientist can more or less consciously tread a path of inner salvation. Slowly then develop inner images, fantasies or ideas, compensatory to the external situation, which indicate the possibility of a mutual approach of poles in the pairs of opposites. ... As against the strict division of the activities of the human spirit into separate departments since the seventeenth century, I still regard the conceptual aim of overcoming the contrasts, an aim which includes a synthesis embracing the rational understanding as well as the mystic experience of one-ness, as the expressed or unspoken mythos of our own present age." ([P] 16. pp. 147-148).

*Acknowledgments*

I thank Stanley Deser for his critical comment on the first version of this paper.

Work supported in part by Grant DFNI T02/6 of the Bulgarian Science Foundation.